\documentclass[aps,prd,twocolumn,showpacs,groupedaddress]{revtex4}
\usepackage{graphicx}
\newcommand{\be}{\begin{equation}}
\newcommand{\ee}{\end{equation}}
\newcommand{\ba}{\begin{eqnarray}}
\newcommand{\ea}{\end{eqnarray}}
\newcommand{\Eq}[1]{Eq.~(\ref{#1})}
\newcommand{\Eqs}[1]{Eqs.~(\ref{#1})}
\newcommand{\rf}[1]{(\ref{#1})}
\newcommand{\Ref}[1]{Ref.~\cite{#1}}

\newcommand{\Sec}[1]{Sec.~\ref{#1}}

\newcommand{\Fig}[1]{Fig.~{\ref{#1}}}

\begin{document}

\title{Confinement Phenomenology in the Bethe-Salpeter Equation} 
\author{M.~Bhagwat}
\author{M.~A.~Pichowsky}
\author{P.~C.~Tandy}
\affiliation{Center for Nuclear Reseach, Department of Physics,
             Kent State University, Kent, Ohio 44242 U.S.A.}
\date{\today}
\begin{abstract}
We consider the solution of the Bethe-Salpeter equation in Euclidean metric
for a $\bar q q$ vector meson in the circumstance where the dressed 
quark propagators have time-like complex conjugate mass poles.   This
approximates features encountered in recent QCD modeling via the 
Dyson-Schwinger equations; the absence of real mass poles simulates
quark confinement.   The analytic continuation in the total momentum 
necessary to reach the mass shell for a meson sufficiently heavier than 
1~GeV leads to the quark poles being 
within the integration domain for two variables in the standard approach.
Through Feynman integral techniques, we show how the analytic continuation
can be implemented in a way suitable for a practical numerical solution.
We show  that the would-be $\bar q q$ width to the meson generated from 
one quark pole is exactly cancelled by the effect of the conjugate
partner pole; the meson mass remains real and there is no
spurious $\bar q q$ production threshold. 
The ladder kernel we employ is consistent with one-loop perturbative QCD
and has a two-parameter infrared structure found to be successful in 
recent studies of the light SU(3) meson sector.
\end{abstract}


%

\maketitle

\section{Introduction \label{Sec:Intro}}

QCD models based on the
Dyson--Schwinger equations [DSEs] provide an excellent tool in the study
of nonperturbative aspects of hadron properties in QCD \cite{Roberts:2000aa}.  
Such models can implement quark and gluon
confinement \cite{Roberts:2000aa,Burden:1992gd,Krein:1992sf,Maris:1995ns}, 
generate dynamical chiral symmetry
breaking \cite{Atkinson:1988mw,Roberts:1990mj}, and maintain Poincar\'e 
covariance.  It is
straightforward to implement the correct one-loop renormalization group
behavior of QCD \cite{Maris:1997tm}, and obtain agreement with perturbation
theory in the ultraviolet region.  Provided that the relevant Ward
identities are preserved in the truncation of the DSEs, the
corresponding currents are conserved.  Axial current conservation
induces the Goldstone nature of the pions and kaons~\cite{Maris:1998hd};
electromagnetic current conservation produces the correct hadronic
charge without fine-tuning~\cite{Oettel:1999gc}.   

Previous work~\cite{Maris:1997tm,Maris:1999nt} has shown this model to 
provide an efficient description of the light-quark pseudoscalar and vector 
mesons via two infrared parameters within the rainbow 
truncation of the DSE for solution of the dressed quark propagators 
coupled to the ladder approximation for the Bethe-Salpeter equation [BSE].
Furthermore, in impulse 
approximation, the elastic charge form factors 
of the pseudoscalars~\cite{Maris:2000sk} and the electroweak transition 
form factors of the pseudoscalars and vectors~\cite{Maris:2002mz,Ji:2001pj} 
are in excellent agreement with data.   As a collorary, the strong decays 
of the vector mesons are well-described in impulse approximation 
without parameter adjustment~\cite{Jarecke:2002xd}.  

Nonperturbative solutions of the DSEs are almost always implemented in
the Euclidean metric for technical reasons and this work is no exception.
A resulting complication is that the solution of the BSE for meson
bound states requires an analytic continuation in the meson momentum to
reach the on-mass-shell point $P^2=-M^2$.   This causes the quark $p^2$ 
in the BSE to vary throughout  
a domain of the complex plane bounded by a parabola which widens in 
proportion to the mass $M$ of the meson.     
For low mass mesons, such as the $\pi$ and $K$ mesons, this
excursion into the complex plane is simple to handle. 
Although the difficulties encountered in studies of ground state vector
mesons are greater, they too may be overcome directly~\cite{Maris:1999nt}.    
For mesons with masses greater than about $1~{\rm GeV}$, 
these excursions into the complex plane are deep enough to encounter
singularities in the rainbow DSE solutions for the quark propagators.    
As a result, little is known about the implementation of the BSE in this
situation or even whether a solution is well-defined.
For this reason, DSE-BSE model studies of mesons have for the most
part been restricted to light mesons.  An exception is the
study of Ref.~\cite{Jain:1993qh} in which the Ansatz adopted for the quark
propagators in the non-Euclidean domain precluded singular behavior.
  
It has long been known that the rainbow approximation to the fermion
DSE produces complex conjugate singularities on the timelike half of the
complex-momentum plane (${\Re}(p^2)<0$) in
QED~\cite{Fukuda:1976zb,Atkinson:1979tk}.   
The phenomenon  has been studied in 
detail within QED~\cite{Maris:1992cb,Maris:1994ux} and in an
Abelian-like model of QCD~\cite{Stainsby:1992hy}.   In these early studies
the singularities were attributed to artifacts of the rainbow
approximation.   
A connection between the generation of complex conjugate 
singularities and the possible absence of real mass poles was
noticed in a  QED study~\cite{Fukuda:1976zb}.  Later studies within
QCD~\cite{Burden:1992gd,Krein:1992sf,Gribov:1999ui} raised suggestions 
that complex
singularities may not be artifacts of the approximation at all, but rather
may be a property of the full theory, possibly related to confinement and
the absence of real quark mass poles.
Beyond the rainbow approximation, very little is known about DSE solutions 
and their singularity structure.  
These questions and the influence of approximations have yet to be
completely resolved.
However they are being addressed; for example, in one simple model
calculation, the inclusion of the full dressing of the quark-gluon vertex
in the DSE leads to the disappearence of all quark singularities in the
complex plane save essential singularities at
infinity~\cite{Burden:1992gd}.     

It has been well-established by Dyson-Schwinger studies 
that in order to obtain an acceptable amount of dynamical chiral symmetry
breaking in agreement with empirical observations, the effective rainbow
DSE  kernel must have a significant infrared strength.   
This in turn tends to move mass poles away from the real timelike $p^2$ axis,
at least for a domain characterized by several GeV.  
This suggests that quark propagators may be successfully modeled in terms
of parametrizations based on entire functions~\cite{Roberts:1996hh} or
at least based on functions with singularities away from the real axis
such as complex-conjugate poles.

If such an approach to modeling QCD is to be extended to include 
$\bar{q}q$ and $\bar{q}Q$ meson spectra beyond 1 or 2~GeV, 
then it is necessary to determine how solutions of the BSE are defined
when complex singularities in the quark propagators may be encountered.
This issue is addressed in the following using a simple parmetrization of
the quark propagator. 
Although the method developed herein is first applied to the BSE, it has
applications beyond meson spectroscopy.   
For example, in production processes at moderate momentum transfers, 
the use of complex mass pole representations for confined quarks 
has been suggested as a means to handle the sizable timelike momentum 
carried by the quarks in these processes~\cite{Ahlig:2000qu}.   
One deals with these issues also in applications of the DSE-BSE approach to 
electromagnetic elastic~\cite{Maris:2000sk} and transition form
factors~\cite{Maris:2002mz,Ji:2001pj} of light mesons at high-momentum 
transfer. 

To simplify the analysis, we adopt a parameterization of the 
dressed quark propagator in terms of complex conjugate mass poles.  
In the standard approach to the meson BSE, two of the four integration 
variables are integrated numerically and it is during the integration of
these variables that poles may be encountered as a result of analytic
continuation. 
In this article, a clear implementation, based on Feynman integral
techniques, is introduced to map the four-dimensional integration in the
BSE onto a one-dimensional domain where numerical methods of evaluating
Cauchy integrals are well-established. 
We find that when a complex conjugate mass pole parametrization of the quark
propagators is employed, the would-be $\bar{q} q$ decay width of the meson  
generated from a particular mass pole is exactly cancelled by the 
contribution of the conjugate partner pole.
As a result, the meson mass $M$ remains real valued, even above the 
pseudo-threshold \mbox{$M^2 > \Re(4m_1^2)$} where 
$m_1$ is the lowest complex mass in the propagator representation.

The primary limitation of the method developed here is that it relies
on an explicit parameterization of dressed quark propagators  in terms
of pairs of complex conjugate mass poles.   
However the limitation is offset by the finding that such a 
representation can provide excellent fits to both DSE solutions (as is 
shown in \Sec{Sec:ModelCalcs}) and also to
recent lattice-QCD data~\cite{detmold_privcom02}.  Of course, use of such 
a phenomenological representation will not exactly preserve the dynamical 
relation between vertices and propagators necessary to satisfy the Ward
Takahashi identities of QCD.   In particular, without the axial 
vector Ward Takahashi identity, the chiral-limit pseudoscalar meson
solutions will not obey Goldstone's theorem~\cite{Maris:1998hd}.
As a result, the toy model introduced in \Sec{Sec:ModelCalcs} would not
be recommended for studies of light pseudoscalar mesons; 
chiral-symmetry-preserving DSE approaches have already been
well-established~\cite{Maris:1997tm,Maris:1999nt} for that sector. 
Rather, the present work explores techniques for extending the present
DSE-based approach to mesons with masses above 1~GeV where one
expects to encounter complex conjugate singularities.

The article is organized as follows.
In Sec. II, our notation for the ladder BSE for meson
states is introduced, as is the complex mass pole representation of the
quark propagators which will be employed.    
In Sec. III,  a simplified BSE integral equation is used to
illustrate the technique for  analytic continuation of the BSE.
The fact that the resulting bound state mass will be real is demonstrated
explicitly. 
In Sec. IV, the method is extended to the general BSE for a vector meson.
A simple model for the ladder BSE kernel is introduced in Sec.~V 
and numerical results are presented.
A discussion of this approach and outlook is presented in Sec.~VI.


%
\section{The Bethe-Salpeter Equation \label{Sec:BSE} }

In the following the Euclidean metric is employed.
The scalar product of two four vectors is then 
\mbox{$a\cdot b=\sum_{\mu=1}^4 a_\mu b_\mu $}, and 
\mbox{$a^2 > 0$} for a spacelike 4-vector $a_\mu$.
The Dirac $\gamma$-matrices are Hermitian 
\mbox{$\gamma_\mu^\dagger = \gamma_\mu$}  and obey the anti-commutation
relation \mbox{$\{\gamma_\mu,\gamma_\nu\} = 2\delta_{\mu\nu}$}. 
The dressed quark propagator $S(p)$ and meson Bethe-Salpeter (BS)
amplitude $\Gamma(p;P)$  are solutions of the renormalized 
DSE,
\begin{eqnarray}
S(p)^{-1}\!\!&=&\!Z_2 \, i\,/\!\!\!p + Z_4 \, m_{q}(\mu) \; 
\nonumber\\ &&
    +Z_1 \!\int^\Lambda_q \!\!\! g^2D_{\mu\nu}(p-q) \,
\frac{\lambda^i}{2}\gamma_\mu \, 
S(q) \, \Gamma^i_\nu(q,p),
\label{quarkdse}
\end{eqnarray}
and the BSE,
\begin{eqnarray}
\Gamma(p;P) \!&=&\!\! \int^\Lambda_q \!\! K(p,q;P)  \odot
      \big( S(q_+) \Gamma(q;P)S(q_-) \big), \;\;
\label{bse}
\end{eqnarray}
Here $D_{\mu\nu}(k)$ is the renormalized dressed-gluon propagator,
$\Gamma^i_\nu(q,p)$ is the renormalized dressed quark-gluon vertex,
$\Gamma(p;P)$ is the BS amplitude for a quark-antiquark bound-state
meson, and $K(q,p;P)$ is the renormalized two-particle irreducible 
$q\bar{q}$ scattering kernel.   We consider equal mass quarks with
momenta $q_\pm = q \pm P/2$ where $q$ is the
quark-antiquark relative momentum, and \mbox{$P= q_+ - q_-$} is
the meson momentum  which satisfies $P^2 = -M^2$ where
$M$ is the meson mass. 

In \Eq{bse} the double contraction of Dirac indices has been denoted by
$\odot$.  
If the Dirac indices of the elements were reinstated in the BSE and
written as parenthetic Roman letters $(a, b, \ldots)$, then \Eq{bse} would
appear as  
\ba
\Gamma_{(ab)}(p;P) \!&=&\!\! \int^\Lambda_q \!\! 
     K_{(ab;cd)}(p,q;P)  S_{(de)}(q_+) 
\nonumber \\ & & \times
 \Gamma_{(ef)}(q;P)S_{(fc)}(q_-)~~.
\ea
Use of the double contraction operator $\odot$ allows us to suppress Dirac
indices, and simplify the forms of the equations appearing herein. 

In \Eqs{quarkdse} and \rf{bse} 
\mbox{$\int^\Lambda_k \equiv \int^\Lambda d^4 k/(2\pi)^4$}
denotes a translationally invariant ultraviolet regularization
of the momentum space integral with mass-scale $\Lambda$.
Lorentz covariance entails that the solution of Eq.~(\ref{quarkdse}) 
is of the form \mbox{$S(p)^{-1} = i /\!\!\! p A(p^2) +$} \mbox{$B(p^2)$}.
The Lorenz scalars $A(p^2)$ and $B(p^2)$ are renormalized at spacelike
$p^2=\mu^2$ such that \mbox{$A(\mu^2)=1$} and \mbox{$B(\mu^2)=m_q(\mu)$}, 
where $m_q(\mu)$ is the renormalized current quark mass. 
After renormalization, one is free to remove the regularization by taking
the limit \mbox{$\Lambda \to \infty$}.    

Recently, a successful model has been developed that provides an excellent
description of the masses, decays and other properties of the light
pseudoscalar and vector mesons~\cite{Maris:1997tm,Maris:1999nt}. 
The model consists of the rainbow-ladder truncations for the quark DSE
(\ref{quarkdse}) and meson BSE (\ref{bse}) and the use of 
an effective $\bar q q$ interaction constrained to coincide
with  perturbative QCD in the ultraviolet domain and containing 
a phenomenological infrared behavior.
In this particular model, the DSE kernel in rainbow truncation is 
\begin{equation}
\label{ourDSEansatz}
Z_1 g^2 D_{\mu \nu}(k) \Gamma^i_\nu(q,p) \rightarrow
 4\pi \alpha_{\rm eff}(k^2)\, D_{\mu\nu}^{\rm free}(k)\, \gamma_\nu
          \frac{\lambda^i}{2} 
 \,,
\end{equation}
where $D_{\mu\nu}^{\rm free}(k=p-q)$ is the free gluon propagator 
in Landau gauge, and $\alpha_{\rm eff}(k^2)$ is the effective running
coupling.  The ladder truncation of the BSE kernel is 
\begin{equation}
\label{ourBSEansatz}
        K(p,q;P) \to
        -4\pi \alpha_{\rm eff}(k^2)\, D_{\mu\nu}^{\rm free}(k)
        {\frac{\lambda^i}{2}}\gamma_\mu \otimes
        {\frac{\lambda^i}{2}}\gamma_\nu \,,
\end{equation}
where \mbox{$k=p-q$} and $\otimes$ denotes the direct (or Cartesian)
product of  matrices.  
This truncation scheme is self-consistent in that it ensures that 
the dressing of the quark-antiquark vector and axial-vector vertices
generated by Eqs.~(\ref{ourDSEansatz}) and (\ref{ourBSEansatz}) satisfy 
their respective Ward-Takahashi identities.   
This feature is important since the axial-vector Ward-Takahashi
identity guarantees that, in the chiral limit of zero current quark mass,
the ground state pseudoscalar meson bound states are the massless
Goldstone bosons arising from dynamical chiral symmetry breaking 
in accordance with the Goldstone's theorem~\cite{Maris:1997tm,Maris:1998hd}.
The vector Ward-Takahashi identity guarantees electromagnetic
current conservation for mesons and
nucleons if the impulse approximation is used to describe the current
in terms of quarks~\cite{Maris:2000sk,Oettel:1999gc}.    
The rainbow-ladder truncation is particularly suitable for the flavor octets
of light pseudoscalar and vector mesons where higher-order contributions
to the quark-gluon skeleton-graph expansion have significant
cancellations~\cite{Bender:1996bb,Bender:2002as} and so may be safely
neglected. 

With the ladder truncation from \Eq{ourBSEansatz}, the BSE for a light-quark
vector meson is  
\begin{eqnarray}
&&\lambda(P^2) \Gamma_\rho(p;P) = \nonumber \\ 
&&\int^\Lambda_q  
    K^L_{\mu \nu}(k)\, \gamma_\mu\, S(q_+)\, 
           \Gamma_\rho(q;P)\, S(q_-)\, \gamma_\nu~,
\nonumber\\ {}
\label{BSE_ladder}
\end{eqnarray}
with \mbox{$k=p-q$} and 
\be
K^L_{\mu \nu}(k) = - \frac{16 \pi}{3}\, 
               \alpha_{\rm eff}(k^2)\, D_{\mu\nu}^{\rm free}(k)~.
\ee
Here $\Gamma_\rho$ is a transverse four vector.   We will
refer to the vector meson under study here as the $\rho$ meson 
although, due to a number of simplifications made for illustration purposes,
the particular model solutions considered here should not be taken as 
physical representations of the $\rho$ meson.

A linear eigenvalue $\lambda(P^2)$ has been introduced into \Eq{BSE_ladder}
so that it will yield solutions over a continuous range of $P^2$. 
A physical bound state of mass $M_{\rho}$ corresponds to an eigenvalue of 
\mbox{$\lambda(P^2=-M_{\rho}^2) = 1$}.  
To find such a solution, it is clear that
\Eq{BSE_ladder} must be analytically continued from Euclidean space where
\mbox{$P^2>0$} to Minkowski space where \mbox{$P^2<0$} and the 
on-mass-shell condition $P^2 = - M_{\rho}^2$ may be realized. 
In terms of the Euclidean four momentum $P_{\mu}$, such an analytic
continuation would correspond to \mbox{$P_4 =$} 
\mbox{$i\sqrt{\vec{P}^2+M_\rho^2}$}, from which it follows that quark
momenta \mbox{$q_\pm^2  = q^2$} 
\mbox{$ - M_\rho^2/4 \pm i\, \sqrt{q^2}\, M_\rho\, z$}
that enter the BSE are complex functions of the integration variables 
$q^2>0$ and the direction cosine $-1 \leq z \leq +1$.
Therefore, in order to carryout such an analytic continuation of the BSE
\rf{BSE_ladder}, the dressed quark propagators $S(q_{\pm}^2)$ must be known
in the complex-momentum plane within a parabolic region having the
negative real (timelike) point  $-M_{\rho}^2/4$ as the apex and extending
symmetrically about the real axis.

Studies within the rainbow-ladder truncation~\cite{Maris:1999nt}
find that the non-analytic points of $u$- and $d$-quark propagators 
that occur nearest to the origin $q^2=0$ are located at 
\mbox{$q^2 = -0.207 \pm  i 0.331~{\rm GeV}^2$} 
while those of the $s$-quark propagator are found at
\mbox{$q^2 = -0.376 \pm i 0.602~{\rm GeV}^2$}~\cite{Jarecke:2002xd}.
This finding also agrees with qualitative observations from an earlier
study~\cite{Frank:1996uk}.   From the positions of these non-analytic points 
one can define a critical mass $M_{c} = \sqrt{ - P^2 }>0$ above which
these singular points enter the domain of integration for the BSE.
For the model of \Ref{Maris:1999nt}, these critical masses are
$M_{c}=1.09$~GeV for $\bar{u} u$ mesons and $M_{c}=1.47$~GeV for 
$\bar{s} s$ mesons.
Clearly, the non-analytic points lie outside the domains of integration
for ladder BSE calculations of ground state pseudoscalar and
vector mesons, but may lie within the integration domain for heavier
mesons.    
This observation provides one of the main reasons that many BSE studies of
mesons focus primarily on light meson states.

The appearence of poles in the domain of integration of the BSE raises two
questions: 1) does the analytic continuation of the BSE  to timelike 
momenta \mbox{$P^2<0$} yield 
physically meaningful quantities, and if so, 2) by what methods does one 
obtain the solutions of this singular integral equation?  
The goal of this article is to explore these questions.
To facilitate an initial investigation, 
it is convenient to adopt a parametrization of the quark propagator 
$S(p)$ as a sum of $N$ pairs of complex conjugate mass poles, 
\begin{equation}
S(p) = \sum_{n=1}^{N} \left\{ \frac{z_n}{ i /\!\!\! p + m_n}
+ \frac{z_n^*}{ i /\!\!\! p + m_n^*} \right\}~~,
\label{pole_rep}
\end{equation}
where $m_n$ are complex-valued mass scales and $z_n$ are complex
coefficients.  Comparison with the general form 
\mbox{$S(p)^{-1} = i /\!\!\! p A(p^2) +$} \mbox{$B(p^2)$}, allows 
identification of the amplitudes $A(p^2)$, $B(p^2)$, 
and the quark running-mass \mbox{$M(p^2) = B(p^2)/A(p^2)$}.   
Equivalently, comparison with the general form 
\mbox{$S(p) = -i /\!\!\! p \sigma_V(p^2) +$} \mbox{$\sigma_S(p^2)$}
allows identification of the scalar and vector amplitudes
of the propagator, $\sigma_{S}(p^2)$ and $\sigma_{V}(p^2)$
respectively.   These are needed for the BSE. 
For example, the quark propagator of Eq.~(\ref{pole_rep}) has a
scalar amplitude of the form
\begin{equation}
\sigma_S(p^2) = \sum_{n=1}^{N} \left\{ \frac{z_n\, m_n}{ p^2 + m_n^2}
+ \frac{z_n^*\, m_n^*}{ p^2 + {m_n^*}^2} \right\}~~,
\label{ss_pole_rep}
\end{equation}
and a vector amplitude $\sigma_V(p^2)$ which is obtained by removing each
of the $m_n$ and $m_n^*$ from the numerators.   

It is shown in \Sec{Sec:ModelCalcs} that this form of quark propagator
with \mbox{$N = 3$} provides an excellent fit to DSE solutions that
employ realistic effective $q\bar{q}$ interactions.
Furthermore, this form has been used recently to provide a parametric fit
of the chiral-quark propagator~\cite{detmold_privcom02} obtained from
lattice-QCD simulations\cite{Bowman:2002bm,Bowman_privCom02}.


%
\section{Analytic Continuation:  A Simple Example \label{Sec:Simple}}

A general study of the analytic continuation of the BSE is quite involved
and complicated by the Dirac algebra necessary to describe quark-antiquark
states.   
It is prudent to first proceed by simplifying several of these complications
and focus on the issue of analytic continuation of the eigenvalue
$\lambda(P^2)$ and the associated integrations over the complex-conjugate
poles in the quark propagator. 
These restrictions are lifted in Sec.~\ref{Sec:General} where the approach
introduced here is extended to the general BSE.

In the following, assume that the quark propagator is
well-represented by a single pair of complex conjugate poles (that is, 
\mbox{$N=1$} in Eq.~(\ref{pole_rep})). Furthermore, assume that the
vector-meson BS amplitude may be represented by the single Dirac amplitude 
$\Gamma_\rho(p;P) = (\delta_{\rho\alpha}-P_{\rho}P_{\alpha}/P^2)
\gamma_\alpha \, V(q;P)$, where $P_\rho$ is the four momentum of the 
bound-state vector meson.  After these simplifications, the BSE 
(\ref{BSE_ladder}) may be converted to a scalar eigenvalue equation for the 
amplitude $V(q;P)$ by multiplication from the left by $\gamma_\rho$ 
followed by a trace over the Dirac indices to obtain  
\begin{eqnarray}
&&\lambda(P^2) V(p;P) = \nonumber \\ 
&&\sum_{a,b=\pm 1}  \int^\Lambda_q \!\! \frac{f_{ab}(p,q,P)\; V(q;P)}
{( q_+^2 + \mu^2 +i a \Delta^2)\, ( q_-^2 +  \mu^2 + i b
\Delta^2)}.~~
\label{BSE_traced}
\end{eqnarray}
Observe that the kernel of this equation is bilinear in the propagator
amplitudes  $\sigma_{S}(q_\pm^2)$ and $\sigma_{V}(q_\pm^2)$ but only 
the denominators of these functions are displayed explicity in
Eq.~(\ref{BSE_traced}).
The remaining factors from the quark propagators and effective $q\bar{q}$
interaction have been collected into the scalar function $f_{ab}$ whose
detailed form is unimportant to the present discussion but may be derived
following the discussion of Sec.~\ref{Sec:General}.
In addition, the denominators have been written in terms of 
$\mu^2=m_R^2-m_I^2$ and $\Delta^2=2m_R m_I$, where 
\mbox{$m=m_R + i  m_I$} is the complex mass scale parameter in the quark
propagator in \Eq{ss_pole_rep}.    
This form allows a parametrization of the square of the quark mass $m^2$
and its complex conjugate $m^{2 *}$ in terms of the values $a = \pm 1$.

In principle, the $q$-dependence of the integrand  in \Eq{BSE_traced} is
only known once the solution  $V(q;P)$ has been obtained.  
For the present, consider the artificial situation in which the solution 
of the BSE is known to be of the form \mbox{$V = g\, \phi(q^2)\; U(z_q)$} 
where $z_q = q\cdot P / \sqrt{q^2 P^2}$ and $\phi(q^2)$ and $U(z)$ are known 
functions.    
In this case, the constant $g$ may be eliminated from the BSE and
following an integration over $d^4 p$ one obtains, 
\be
\lambda(P^2) = \sum_{a,b=\pm 1} \; \lambda_{ab}(P^2),
\ee
and 
\be
\lambda_{ab}(P^2) = \int\!\!\! \frac{d^4 q}{(2\pi)^4} \frac{F_{ab}(q,P)} 
    {( q_+^2 + \mu^2 +i a \Delta^2)\, ( q_-^2 +  \mu^2 + i b
\Delta^2)}~~, 
\label{BSE_integral}
\ee
where  $F_{ab}(q,P)$ is a non-singular, Lorentz invariant function of $q$
and $P$.  
The objective is then to evaluate the four integrals $\lambda_{ab}(P^2)$ 
and show how they may be analytical continued to negative values of
\mbox{$P^2$}.   
This is carried out explicitly in the remainder of this section.

When $P^2$ is continued to negative values, the quark momenta in 
\Eq{BSE_integral} satisfy \mbox{$q_{\pm}^2 = (q_{\mp}^2)^{*}$}. 
The two factors in the denominator of
$\lambda_{a,-a}(P^2)$ (that is, $a=-b$ in \Eq{BSE_integral}) will thus 
be complex conjugates.   As a result, for $P^2$ sufficiently 
timelike, there will always be some value of the integration  
variable $q_{\mu}$ for which both factors in the denominator of
$\lambda_{a,-a}(P^2)$ vanish simultaneously.    The result is a 
non-analytic point in $\lambda_{+-}(P^2)$ and $\lambda_{-+}(P^2)$. 
Such behavior would usually be indentified with the decay threshold for
the vector meson into an asymptotically-free quark and antiquark pair.
However, in the following, it is shown that when the integrals in
\Eq{BSE_integral} are  added together the imaginary parts cancel
exactly thereby eliminating the possibility of quark-antiquark
production thresholds.
In contrast to this, the two factors in the denominator of 
$\lambda_{aa}(P^2)$ never vanish simultaneously and consequently there
are no corresponding non-analytic points in either $\lambda_{++}(P^2)$
or $\lambda_{--}(P^2)$.   

To analyze the behavior in detail, it is advantageous to restructure
the integral in
\Eq{BSE_integral} so that the analytic continuation from
the  Euclidean domain  \mbox{$P^2>0$} to the timelike domain
\mbox{$P^2<0$} is straightforward. 
For values of \mbox{$P^2>0$}, one may combine the two factors in the
denominator using the Feynman parametrization,
\be
\frac{1}{A B} = \int_{0}^{1} \!d\alpha d\beta \;
\frac{\delta(\alpha + \beta - 1)}{ (\alpha A + \beta B)^2 }
\ee
and obtain
\ba
\lambda_{ab}(P^2) \!\! &=& \!\!\!
\int_{-1}^{+1} \!\! \frac{dy}{2} \int \!\!\! \frac{d^4 q}{(2\pi)^4} 
\; \frac{F_{ab}(q,P)}{D(q,P,y)^2}~,
\label{SimpleGeneralG1}
\ea
where \mbox{$y=\alpha - \beta$}, and 
\be
D(q,P,y) = y ( q \cdot P + i \frac{a\!-\!b}{2} \Delta^2 ) 
  + ( q^2 + \frac{P^2}{4} + \mu^2 + i \frac{a\!+\!b}{2} \Delta^2 )~.
\ee  
One then eliminates the term linear in $q$ in the denominator by a 
shift of the origin \mbox{$q_\nu \rightarrow q_\nu - yP_\nu/2$}.  
The resulting denominator depends on $P^2$, $q^2$, and $y$ but not on  
$q\cdot P$.   After this shift, the function $F_{ab}(q,P)$ in the numerator
still depends $q \cdot P$.   All other variables save $q^2$ and $y$ can 
be integrated out yielding, 
\ba
\lambda_{ab}(P^2) &=& \!\! 
\int_{-1}^{+1} \!\!\! dy \,\int_0^\infty \!\!\! dq^2 
\; \frac{q^2\, \tilde{F}_{ab}(q^2,y;P^2) }{ [q^2 - q^2_{ab}(P^2,y)]^2}, 
\label{SimpleGeneralG2}
\ea
where 
\be
q^2_{ab}(P^2,y) = - \frac{P^2}{4} (1 - y^2) - \mu^2
         -\frac{i}{2} 
         \left(y (a-b) + a+b \right) \Delta^2,
\label{Simpleqq0}
\ee
is independent of $q^2$, and 
\be
\tilde{F}_{ab}(q^2,y;P^2)\! = \!
\frac{1}{16\pi^3} 
\! \int_{-1}^{+1} \!\!\!\! dz_q \, \sqrt{1-z_q^2} \;
F_{ab}(q-\frac{y}{2}P,P)~.
\label{Ftilde}
\ee
Here $z_q$ is the direction cosine defined by 
\mbox{$q \cdot P = \sqrt{q^2 P^2}\, z_q$}.
The singular behavior due to the propagators is now
a single second-order pole in the variable $q^2$.  
Its location $q^2_{ab}(P^2,y)$ is given in \Eq{Simpleqq0} and is not in
the path of integration for spacelike \mbox{$P^2>0$};
the integral is regular in Euclidean space.

\begin{figure}[tb]
\includegraphics[width=3.20in]{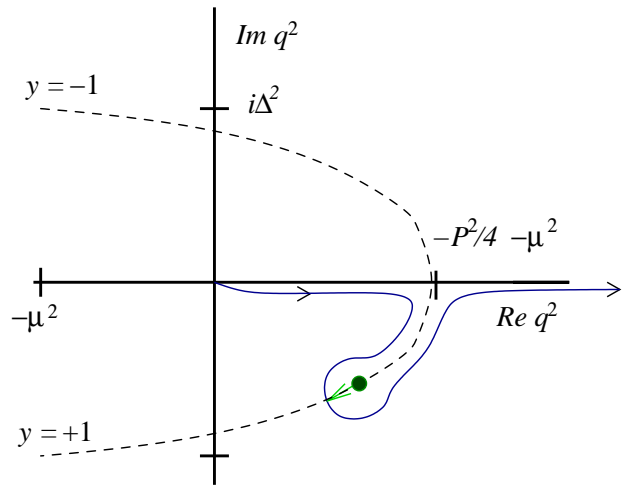}
\caption{The trajectory of pole position $q^2_{+-}(y)$ given by
\Eq{Simpleqq0} is shown as the dashed curve.   The solid curve is the 
contour required to implement analytic continuation. 
\label{Fig:PolePos1} }
\end{figure}

The $y$ integration is performed after the integration over $q^2$.
Hence, for a given value of $y$, the $q^2$-integration domain is the
positive  real axis and the integrand has a pole in the complex $q^2$
plane  whose location $q^2_{ab}(P^2,y)$ moves  with $P^2$.   
The location has a negative real part for \mbox{$P^2>0$}
and remains negative as $P^2$ passes through zero and becomes moderately
timelike.  Only when \mbox{$P^2 < -4\mu^2$} will the real part of
the pole location become positive and have a chance of impinging upon
the domain of the $q^2$-integration.
For that to happen, the imaginary part of $q^2_{ab}(y)$ would have to
vanish.   As anticipated above, \Eq{Simpleqq0} shows that this never 
happens when \mbox{$a=b$};  for
$\lambda_{++}(P^2)$ and $\lambda_{--}(P^2)$ the two factors in the
denominator of \Eq{BSE_integral} are never  zero simultaneously.
It is therefore trivial to carryout these integrations.
However, for \mbox{$a=-b$}, the imaginary part of $q^2_{ab}(y)$ can become
zero when the Feynman variable \mbox{$y=0$} and care must be taken to 
properly handle such singular integrals. 

As a specific example, the trajectory of the pole location $q^2_{+-}(y)$
is shown in \Fig{Fig:PolePos1}.  It begins at $(-\mu^2, \Delta^2)$ when
$y=-1$, moves along the dashed
curve as $y$ increases, crossing the real axis when $y=0$, and terminates at 
$(-\mu^2, -\Delta^2)$ when $y=1$.   
For \mbox{$y<0$},  the calculation of the integral is straightforward.
However for \mbox{$y>0$}, the pole in $q^2$ has crossed the real axis 
and the analytic continuation of the integral $\lambda_{+-}(P^2)$ is
defined by continuous deformation of the contour in a manner that ensures
the pole does not cross the contour of integration.   
This is illustrated in \Fig{Fig:PolePos1}, for the deformed contour
corresponding to $y\approx +1/2$.   
Deformation of a integration contour in this manner is equivalent to
adding the pole residue  to the result of 
integration along the real $q^2$ axis.  If the residue term were 
neglected, one would obtain a  discontinuous result due 
to the crossing of a branch cut onto the wrong sheet.
 
For the integral $\lambda_{-+}(P^2)$, the pole trajectory traces out the same
path as in \Fig{Fig:PolePos1} but in the opposite direction; that is, 
\mbox{$y>0$} corresponds to the upper half $q^2$-plane and the pole
residue will be of opposite sign from $\lambda_{+-}(P^2)$.   
After analytic continuation to \mbox{$P^2<0$}, the results for the four
integrals can be expressed as
\ba
\lambda_{ab}(P^2)=&& \int_{-1}^{+1} \!\!dy \,\int_0^\infty \!\! ds 
  \frac{s\, \tilde{F}_{ab}(s,y;P^2) }{ [s - q^2_{ab}(y,P^2)]^2}
\nonumber \\
&&+ \frac{a-b}{2}\; 2\pi i \; \Theta(-\frac{P^2}{4}-\mu^2)
\nonumber \\
&&\times  \int_0^1 \!\! dy 
\frac{d}{ds} \big(s \; \tilde{F}_{ab}(s,y;P^2)\big)\bigg|_{s=q^2_{ab}}~, 
\label{finalint}
\ea
where $\Theta(x)$ is the Heaviside step function. 

In the timelike region, 
inspection of \Eq{finalint} shows that individual integrals  
$\lambda_{ab}(P^2)$ are complex valued and satisfy
\ba
   [\lambda_{a,b}(P^2)]^* &=& \lambda_{-a,-b}(P^2)~.  
\label{Gsymm}
\ea
The imaginary parts cancel in pairs thereby producing a {\em real} result
for the eigenvalue $\lambda(P^2) = \sum_{a,b=\pm} \lambda_{ab}(P^2)$.   
Notice that this does not mean that one may neglect the addition of
residue
terms required by the analytic continuation of the integrals.
Rather, when \mbox{$\Delta^2 >0$} the residue (second) term of
\Eq{finalint} is not purely imaginary and the sum of the real parts of the
residues are required for the final result for $\lambda(P^2)$.
Of course, these residue terms appear only for $P^2$ more timelike than
the pseudo-threshold point \mbox{$P^2 = -4 \mu^2$}, that is, 
for bound state masses \mbox{$M_{\rho}> 2 \mu$}. 

Before extension of this treatment to the general problem, consider 
the connection
between the above discussion of analytic continuation and the usual case
encountered  above physical two-body thresholds.
In particular, consider the limiting case of infinitesimal $\Delta$ 
for the single integral
$\lambda_{--}(P^2)$.  This corresponds to the standard case of employing
elementary propagators with the conventional boundary conditions of a
negative infinitesimal $-i\epsilon$ added to the real mass.  
Above the two-body threshold \mbox{$P^2 < -4 \mu^2$}, there would be
$\bar q q$
production described by the residue term in $\lambda_{--}(P^2)$.
It follows that a bound state calculation, based on such an integral alone,
would result in a complex mass describing an unstable meson with 
a finite $\bar q q$ decay width.   
However, in the present case the presence of conjugate quark poles ensures
that the analytic continuation introduces both positive and negative 
would-be widths that cancel exactly. 
The resulting BSE eigenvalue $\lambda(P^2)$ in
\Eq{BSE_traced} will be purely real and therefore bound state masses defined
by $\lambda(P^2 = -M_{\rho}^2)=1$ will be real and will describe a meson
with no $\bar q q$ decay width.


\section{Generalization to the Bethe-Salpeter Equation
         \label{Sec:General}} 

The technique  described in the previous section is generalized
to allow solutions to the
homogeneous BSE for an arbitrary quark-antiquark bound state.   To be
specific, we consider an equal quark-antiquark vector meson bound state.
One can write the most-general BS amplitude for a massive vector meson as
a sum of eight Dirac covariants and Lorentz-scalar amplitudes
$F_\alpha(q;P)$,   
\begin{eqnarray}
\Gamma_\rho(q;P) &=& 
      \left(\delta_{\rho\mu}-\frac{P_{\rho}P_{\mu}}{P^2} \right)
       \big(
        \gamma_\mu \, F_1 
        + q_\mu\not\!q \, F_2
\nonumber \\   && {}         
        + q_\mu\not\!P \, F_3
        + \gamma_5\epsilon_{\mu\alpha\nu\beta}\gamma_\alpha 
                q_\nu P_\beta \, F_4
\nonumber \\   && {} 
        + q_\mu \, F_5
        + \sigma_{\mu\nu}q_\nu  \, F_6
        + \sigma_{\mu\nu}P_\nu \, F_7
\nonumber \\   && {} 
        + q_\mu \sigma_{\alpha\beta}q_\alpha P_\beta \, F_8 \;\big).
\label{vecBSAform}
\end{eqnarray}
In a more concise form this is
\be 
\Gamma_\rho(q;P) = \sum_{\alpha =1}^8 \; \Lambda^\alpha_\rho(q,P)\, 
                                                  F_\alpha(q;P)~,
\label{BSAcovsum}
\ee
where $\Lambda^\alpha$ are the Dirac covariants in Eq.~(\ref{vecBSAform});
for example, $\Lambda^{\alpha=1}_{\rho} =$
$(\delta_{\rho\mu}-P_{\rho}P_{\mu}/P^2) \, \gamma_{\mu}$.  
When all indices of a product of two Dirac covariants are contracted in the
manner corresponding to the usual trace operation, one obtains 
\ba
{\rm tr}_{\rm D}
\left[ \Lambda_{\mu}^{\alpha}(q,P) \; \Lambda_{\mu}^{\beta}(q,P) \right] 
&=&
 \Lambda_{\mu}^{\alpha}(q,P) \odot \Lambda_{\mu}^{\beta}(q,P),  
   \nonumber \\
&=& t_{\alpha \beta}(q,P), \label{Def:t}  
\ea
where $t_{\beta\alpha}(q,P)$ is a Lorentz scalar function of $q^2$, $P^2$
and $q\cdot P$, and we have used the
double-contraction product $\odot$ notation introduced in \Eq{bse}. 

Of course, one is free to choose any other set of Dirac covariants
$\Lambda^{\alpha}_{\rho}(q,P)$ as long as they are complete.
In particular, one is free to choose a set of covariants for which 
\mbox{$t_{\alpha \beta} \to \delta_{\alpha \beta}$} in which case 
the resulting set of coupled integral equations may be of a simpler form
than those obtained herein.  However, in practice the behavior of such
covariants can be poorly behaved near \mbox{$q=0$}, and can lead to
numerical difficulties~\cite{Maris:1999nt}. 
Therefore, the set of covariants given by
Eq.~(\ref{vecBSAform}) is preferable and the lack of
orthogonality is easily accommodated by $t_{\alpha\beta}(q,P)$.  

In the previous section, it was shown that when the $q^2$ and $q\cdot P$
dependence of the BS amplitude $\Gamma_\mu(q;P)$ is known, the
analytic continuation of the BSE may be carried out in a straight-forward
manner.  
In practice, the functional form of the BS amplitude is
known only {\em after} the solution to the BSE has been obtained. 
The approach outlined in the previous section may be generalized
by expansion of the 
functional dependence of the BS amplitudes $F_{\alpha}(q;P)$ in terms of
complete sets of known functions of $q^2$ and 
\mbox{$z_q= {q} \cdot {P} / \sqrt{q^2 P^2}$}.  
Consider the expansion
\be
F_\alpha(q;P) = \sum_{s,i=0}^{\infty}\, 
       U_i(z_q)\, \phi_s(q^2)\, g^\alpha_{si}(P^2)~,
       \label{ExpandF}
\ee
where $U_i(z_q)$ are Chebyshev polynomials of the second kind 
satisfying the ortho-normality condition
\begin{equation}
\frac{2}{\pi}\int_{-1}^1\,dz\,\sqrt{1-z^2}\,U_i(z)\,U_j(z) = \delta_{ij}~,
\end{equation}
and $\phi_s(q^2)$ are a set of ortho-normal functions, that are 
specified later, and that satisfy
\be
\int_{0}^{\infty} \!\!\! dq^2 \; \phi_{s}(q^2) \; 
                                       \phi_{t}(q^2) = \delta_{st}~.
\label{phi-orthog}
\ee
The required number of basis terms will be determined by 
convergence of the BSE solution with respect to the meson observables 
of interest.  
For example, one particular choice of basis $\{\phi_{s}(q^2)\}$ which
proves to be efficient for observables dominated by infrared
physics may be inefficient for observables sensitive to the
ultraviolet.  In the present work, we take convergence of the meson
mass as the criterion.  

For brevity, we define the operator ${\cal O}_{si}(p)$ which 
projects out the expansion coefficients $g^{\alpha}_{si}(P^2)$ from
the BS amplitude in Eq.~(\ref{ExpandF}).  
Using \Eqs{BSAcovsum} and \rf{Def:t}, one finds
\ba
g^{\alpha}_{si}(P^2) \!&=&\! {\cal O}_{si}(p) \; 
            t^{-1}_{\alpha\alpha'}(p,P) \Lambda^{\alpha'}_{\rho}  \odot 
            \Gamma_{\rho}(p;P)
\nonumber \\
               &=& \!\! \int_{0}^{\infty} \!\!\!\! dp^2 \!
           \int_{-1}^{+1} \!\!\!\! dz_p \sqrt{1-z_p^2}
\; U_{i}(z_p) \,\phi_{s}(p^2) \nonumber \\
       && \times t^{-1}_{\alpha \alpha^\prime}(p,P)\, 
                   {\rm tr}_{\rm D}
                    \left[ \Lambda_{\rho}^{\alpha^\prime}(p,P) \; 
  \Gamma_{\rho}(p;P) \right].
\label{LHSproj}
\ea
The set of coefficients $g^{\alpha}_{si}(P^2)$ will be refered to as BS
amplitudes since they are equivalent to the BS amplitude
$\Gamma_{\rho}(p;P)$ in that they contain all the dynamical 
information regarding the solution of the BSE. 
The expansion coefficients $g^\alpha_{si}(P^2)$ are functions of $P^2$
only.    The operator ${\cal O}_{si}(p)$  is an {\em  integral} 
operator that is introduced for brevity and notational convenience. 

When both sides of the BSE \rf{BSE_ladder} are projected according to 
\Eq{LHSproj} one obtains
\ba
  \lambda(P^2) g^{\alpha}_{si}
&=&   {\cal O}_{si}(p) 
   \!\! \int \!\!\! \frac{d^4q}{(2\pi)^4} 
   t^{-1}_{\alpha \alpha'}(p,P) 
   \bigg[ \Lambda_{\rho}^{\alpha'}(p) 
\nonumber \\
&&  \odot K(p,q;P) 
 \odot \big( S(q_+) \Gamma_{\rho}(q;P) S(q_-) \big) \bigg],
   \nonumber  \\
&=& \!\!\! \sum_{AB\beta t j} \! 
    {\cal O}_{si}(p)
     \int \!\!\! \frac{d^4q}{(2\pi)^4} 
    t^{-1}_{\alpha \alpha'}(p,P)
     \bigg[ 
     \Lambda_{\rho}^{\alpha'}(p) 
\nonumber \\ && \!\!\! 
\odot
   K(p,q;P) \odot 
   \big( \widetilde{\Lambda}^{A}(q_+) 
     \Lambda_{\rho}^{\beta}(q;P) 
     \widetilde{\Lambda}^{B}(q_-) \big) \bigg]
 \nonumber \\ && \times
   \sigma_{A}(q^2_{+}) \sigma_{B}(q_{-}^2) 
   \phi_{t}(q^2) U_{j}(z_{q}) g^{\beta}_{tj}, 
     \label{BSE_general}
\ea
where the sums are over $\beta \in \{1,\ldots,8\}$, and 
$t,j \in \{0,1,\ldots\}$, and  $A,B \in \{V,S\}$.
The second equality follows by expanding the BS amplitude
$\Gamma_{\rho}(q;P)$  according to
\Eqs{BSAcovsum} and \rf{ExpandF}. 
In \Eq{BSE_general} another set of Dirac covariants, 
$\widetilde{\Lambda}^{A}(q) = \{ -i \gamma \!\cdot\! q ,\; 1 \}$ with 
$A=V, S$, have been introduced to allow the
propagator representation  
$S(q)= \sum_{A=V,S} \widetilde{\Lambda}^{A}(q) \sigma_{A}(q^2)$. 

At this point, it is helpful to specify a coordinate system.
We take \mbox{$P_{\mu}=$} \mbox{$\sqrt{P^2}(0,0,0,1)$}, \mbox{$p_{\mu}=$}
\mbox{$\sqrt{p^2}(0,0,\sqrt{1-z_p^2},z_p)$},
and the integration momentum  $q_{\mu}$ is represented as
\be
q_{\mu}=\sqrt{q^2}\left(
  \begin{array}{c}
   \cos\phi\sqrt{1-x_{q}^2}\sqrt{1-z_{q}^2} \\
   \sin\phi\sqrt{1-x_{q}^2}\sqrt{1-z_{q}^2} \\
    x_q\sqrt{1-z_q^2} \\ 
    z_q
    \end{array} \right)~.
\ee
The integrand in \Eq{BSE_general} has no dependence upon angle $\phi$; the
only dependence on direction cosine $x_q$ is through the $p \cdot q$
dependence of the BS kernel $K(p,q;P)$.   In \Eq{BSE_general} we may
therefore substitute
\ba
\int d^4q = \pi \int_0^\infty dq^2 \, q^2 
\int_{-1}^1 dz_q \,\sqrt{1-z_q^2} \,\int_{-1}^1 \!\!\! dx_{q}~~.
\label{measure}
\ea
The integration over $x_q$ in \Eq{BSE_general} produces a function 
of $p^2, q^2, z_p, z_q$ and $P^2$ which can be expanded in the basis
functions.  That is, we define a quantity ${\cal V}$ by
\ba
\lefteqn{
\int_{-1}^{+1} \!\!\! dx_{q} \; \; 
     t^{-1}_{\alpha\alpha'}(p,P)  \;  
       \bigg[ \Lambda^{\alpha'}_{\rho}(p) \odot
  K(p,q;P) }
\nonumber \\ && \odot 
  \big( \widetilde{\Lambda}^{A}(q_+) 
     \Lambda_{\rho}^{\beta}(q;P) 
     \widetilde{\Lambda}^{B}(q_-) \big) \bigg]
 \nonumber \\  &=&  \sum_{stij} \phi_{s}(p^2) U_{i}(z_p) \; 
    {\cal V}^{\alpha A;\beta B}_{si;tj}(P^2)
     \; \phi_{t}(q^2) U_{j}(z_q)~. 
\label{geninteraction}
\ea
Upon substituting this back into \Eq{BSE_general} above, one obtains a 
{\em discretized} form of the BSE, 
\ba
\lambda(P^2) \; g_{si}^{\alpha}(P^2) = 
                     {\cal V}^{\alpha A;\beta B}_{si;w\ell}(P^2) 
                \; {\cal G}^{AB}_{w\ell,tj}(P^2) \; g_{tj}^{\beta}(P^2)~,
\label{BSEdiscrete}
\ea
where repeated indices are understood as being summed over.   Here
we have introduced the projection of the product of propagator
amplitudes in the form
\ba
{\cal G}^{AB}_{si,tj}(P^2) \!&=& \!\frac{\pi}{(2\pi)^4} \!\int_0^\infty \!\!
\!\!\!dq^2  q^2 \!  \int_{-1}^1 \!\!\! dz_q \, \sqrt{1-z_q^2} 
\phi_{s}(q^2) U_{i}(z_q)
\nonumber \\ && \times 
\sigma_A(q_+^2) \sigma_B(q_-^2)\; 
\phi_{t}(q^2) U_{j}(z_q)~.
\label{Def:Green}
\ea

Thus far, the BSE has been 
reduced to the discrete eigenvalue problem given in
\Eq{BSEdiscrete}.   Physical solutions are at  
$\lambda(P^2=-M_{\rho}^2)=1$ from which one identifies the existence of
a bound state vector meson of mass $M_{\rho}$.
The only task remaining is the determination of the elements in
\Eq{BSEdiscrete} that make up the kernel of the eigenvalue problem.
These are ${\cal V}$ and ${\cal G}$.   The generalized interaction
${\cal V}$ is determined in terms of the kernel by the projection 
in \Eq{geninteraction}.

Consider the explicit form of ${\cal G}$.  
Clearly there is a strong similarity between the form of 
Eq.~(\ref{Def:Green}) and the simple integral whose analytic continuation
was considered in \Sec{Sec:Simple}.   This is clarified by the expansion
of the quark propagator amplitudes in terms of the mass pole terms
from \Eq{ss_pole_rep} in the form
\be
\sigma_{A}(q^2) = \sum_{n=1}^N \; \sum_{a=\pm 1}\; 
\frac{Z_{n}^{A a}} {q^2 + \mu_n^2 +ia \, \Delta_n^2}~,
\ee
where $Z^{A \;a=+1}_{n} = z_{n}$ or $z_{n} m_{n}$ for $A=V$ or $S$,
respectively, and  $Z^{A \; a=-1}_{n} = z_{n}^*$ 
or $z_{n}^* m_{n}^*$, respectively.   Then, Eq.~(\ref{Def:Green})
can be written as 
\ba
&&{\cal G}^{A B}_{st;ij}(P^2) = \frac{1}{2} \sum_{n,m = 1}^{N} 
\sum_{a,b = \pm 1}\; Z_{n}^{A a} Z_{m}^{B b}
\nonumber \\ &&
\int\!\! \frac{d^4 q}{(2\pi)^4} \; \frac{\phi_s(q^2) U_i(z_q) \, 
                                                \phi_{t}(q^2) U_{j}(z_q)}
{ (q_{-}^2 + \mu_{n}^2 + ia \Delta^2_{n})
      (q_{+}^2 + \mu_{m}^2 + ib \Delta^2_{m}) }.~{}
\label{Green2}
\ea
These integrals in \Eq{Green2} are of the type $\lambda_{ab}(P^2)$
already explored in our discussion of analytic continuation in
\Sec{Sec:Simple}.
The only difference is that the number of integrals that have to
be performed can be large since it depends on the number of basis 
terms for representation of the BS amplitudes via Eq.~(\ref{ExpandF}), 
and the number of 
conjugate pairs of mass poles for representation of the propagators. 

The same methods employed in \Sec{Sec:Simple} may be used here to
calculate the integrals 
${\cal  G}^{AB}_{st;ij}(P^2)$ for all values of $P^2$ while accounting
properly for the analytic continuation.
In particular, the Feynman integral method used here will lead  to a
denominator $[q^2- q^2_{nmab}(P^2,y)]^2$, where the pole location is
\ba
{q}_{nmab}^2(y) &=& -\frac{P^2}{4}(1 - y^2)  
   + y \frac{\mu^2_{n} - \mu^2_{m}}{2}  
   + \frac{\mu^2_{n} + \mu^2_{m}}{2} 
   \nonumber \\ & &
   + i y \frac{a\Delta^2_{n} - b\Delta^2_{m}}{2}  
   + i \frac{a\Delta^2_{n} + b\Delta^2_{m}}{2}~.
   \label{PolePos2}
\ea
The integration over $q^2$ is carried out first, followed by the integration 
over the Feynman parameter $y$, and finally the integration over 
$z_{q}$.  
Integrations over the two remaining angles in \Eq{Green2} are trivial. 
In general, for a quark propagator parametrized in terms of $N$ pairs of
complex-conjugate poles, there are $(2N)^2$ pole trajectories, half of
which impinge on the $q^2$ integration domain and lead to residue 
additions.   The imaginary parts of the integrals in \Eq{Green2} cancel 
in pairs as described in the previous section.

In ladder approximation, the determination of ${\cal V}$ is
straight-forward.  There are never any poles encountered when $P^2$ is
analytically continued to negative values.  Of course, singularities may
be encountered when contributions beyond ladder are maintained in the BS
kernel, and these must be treated in a way analogous to ${\cal G}$.
This possibility is not addressed in the present work.


\section{Model Calculations \label{Sec:ModelCalcs}}
\subsection{Ladder-Rainbow Kernel}

Here we use a particular model for the DSE-BSE ladde-rainbow kernel to 
numerically 
implement the preceeding developments.   The model is specified by the
``effective coupling'' $\alpha_{\rm eff}(k^2)$ and we employ the
Ansatz~\cite{Maris:1999nt}
\begin{eqnarray}
\label{gvk2}
\frac{4\pi \alpha_{\rm eff}(k^2)}{k^2} &=&
        \frac{4\pi^2\, D \,k^2}{\omega^6} \, {\rm e}^{-k^2/\omega^2}
\nonumber \\ && {}
        + \frac{ 4\pi^2\, \gamma_m \; {\cal F}(k^2)}
        {\textstyle{\frac{1}{2}} \ln\left[\tau + 
        \left(1 + k^2/\Lambda_{\rm QCD}^2\right)^2\right]} \;,
\end{eqnarray}
with \mbox{$\gamma_m=12/(33-2N_f)$} and \mbox{${\cal F}(s)=(1 -
\exp\frac{-s}{4 m_t^2})/s$}.  The ultraviolet behavior matches
that of the 1-loop QCD running coupling $\alpha_s(k^2)$.  The resulting
solution of the ladder-rainbow DSE-BSE system of equations in the UV region
generates the correct 1-loop perturbative QCD structure.  The first 
term implements the strong
infrared enhancement in the region \mbox{$0 < k^2 < 1\,{\rm GeV}^2$}
phenomenologically required~\cite{Hawes:1998cw} to produce a realistic
value for the chiral condensate.  With \mbox{$m_t=0.5\,{\rm GeV}$},
\mbox{$\tau={\rm e}^2-1$}, \mbox{$N_f=4$}, \mbox{$\Lambda_{\rm QCD} =
0.234\,{\rm GeV}$}, and a renormalization scale \mbox{$\mu=19\,{\rm
GeV}$}, it has been found that \mbox{$\omega = 0.4\,{\rm GeV}$} and 
\mbox{$D=0.93\,{\rm GeV}^2$} give a good description of
$\langle \bar q q\rangle$, $m_{\pi/K}$ and $f_{\pi}$ with physically 
acceptable current quark masses,~\cite{Maris:1997tm,Maris:1999nt}.
The propagator amplitudes from this rainbow DSE model have recently been
shown~\cite{Jarecke:2002xd} to have the same qualitative features as 
lattice QCD simulations~\cite{Bowman:2002bm,Bowman_privCom02}.

The vector meson masses and electroweak decay constants produced by
this model are in good agreement with experiments~\cite{Maris:1999nt}.
Without any readjustment of the parameters, this model agrees
remarkably well with the most recent Jlab data~\cite{Volmer:2000ek}
for the pion charge form factor $F_\pi(Q^2)$.  Also the kaon charge
radii and electromagnetic form factors are well
described~\cite{Maris:2000sk,Maris:2000wz}.  The strong decays of the
vector mesons into a pair of pseudoscalar mesons are also
well-described within this model~\cite{Maris:2001rq,Jarecke:2002xd}.

\subsection{Results \label{Sec:Results}}

\begin{figure}[ht]
\includegraphics[width=7.5cm, angle=-90]{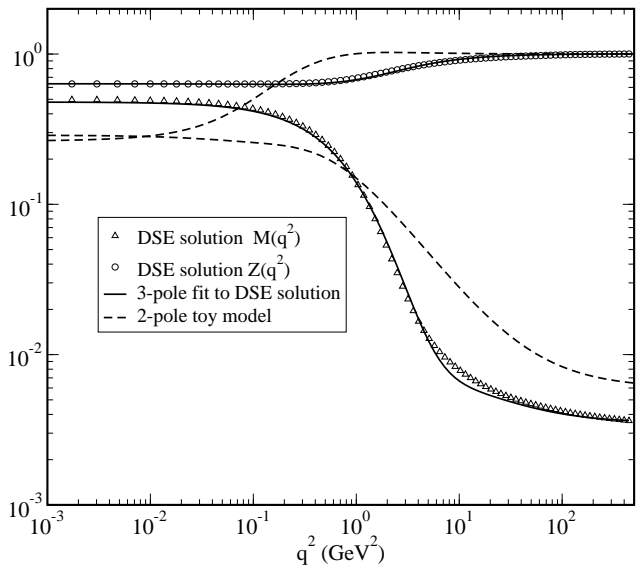}
\caption{
The quark propagator running mass $M(q^2)$ and wave-function
renormalization function $Z(q^2)$.   The DSE 
calculation~\protect\cite{Maris:1999nt} (circles) is compared to 
the fit in terms of three pairs of
complex conjugate mass poles given in \protect\Eq{MTfit}.
The toy model used to test the Bethe-Salpeter method uses two pairs of
poles given in \protect\Eq{toymodel}. 
\label{Fig:MZ}  }
\end{figure}

The DSE solution for the $u/d$ quark propagator from
the model interaction given in \Eq{gvk2} can be well fit with
$N=3$ pairs of complex conjugate poles in the representation 
given in \Eq{pole_rep}.    The result displayed in \Fig{Fig:MZ}
corresponds to the parameter set  
\ba
m_{1}&=& 0.547 + i0.303~{\rm GeV},
\nonumber    \\
z_1&=& 0.200 + i0.475, 
\nonumber    \\
m_{2}&=&-1.262 + i0.570~{\rm GeV},
\nonumber     \\  
z_2 &=& 0.142 + i0.045, 
\nonumber    \\
m_{3}&=& 1.560 + i0.564~{\rm GeV},
\nonumber     \\  
z_3 &=& 0.160 + i0.015~.
\label{MTfit}
\ea
This parameterization was chosen so that $Z(p^2)$ and $M(p^2)$ 
were reproduced well at both $p^2=0$ and $p^2=100~{\rm GeV}^2$
while the main features of the momentum dependence are preserved.
We imposed the constraint that $M(p^2)$ should approach its
UV limit from above.
The pole locations are \mbox{$p^2=-m_i^2$} where 
\mbox{$m_i^2 = $} \mbox{$\mu_i^2 +i\Delta_i^2$}.  Thus we have
\mbox{$\mu_1=$} \mbox{$0.455~{\rm GeV}$}, 
\mbox{$\Delta_1=$} \mbox{$0.576~{\rm GeV}$}, 
\mbox{$\mu_2=$} \mbox{$1.13~{\rm GeV}$},
\mbox{$\Delta_2=$} \mbox{$1.12~{\rm GeV}$}, and
\mbox{$\mu_3=$} \mbox{$1.46~{\rm GeV}$},
\mbox{$\Delta_3=$} \mbox{$1.33~{\rm GeV}$}.

With use of this propagator representation in the BSE, a $u/d$ quark
meson bound state with mass greater than the lowest pseudo-threshold 
$2\mu_1 =$0.910~GeV would be needed to test the method we have described.   
The model-exact $\rho$ meson mass from the  kernel
\Eq{gvk2} is 0.742~GeV~\cite{Maris:1999nt} and is therefore not suitable. 
Heavier $u/d$ mesons have not been carefully studied within this model
interaction due to the very singularity issue we are concerned with.   
In fact the evidence from exploratory studies is that, for example,
$m_{a_1}$ in this ladder-rainbow model is only about 200~MeV above 
$m_\rho$ and also does not provide a clear test.  

To test the present method, we artificially modify the pole 
parameterization of the quark propagator so that the lowest pseudo-threshold 
moves below the vector meson mass, and the latter moves to 0.770~GeV.    
We achieve this with the following parameterization in terms of \mbox{$N=2$}
pairs of conjugate mass poles
\ba
m_{1}&=&-0.400 + i 0.200~{\rm GeV},
\nonumber    \\
z_1&=& 0.15, 
\nonumber    \\
m_{2}&=& 0.550 + i 0.350~{\rm GeV},
\nonumber     \\  
z_2 &=& 0.35 + i0.37~~.  
\label{toymodel}
\ea
The resulting propagator amplitudes are also shown in  \Fig{Fig:MZ}. 
This parameterization corresponds to 
\mbox{$\mu_1=$} \mbox{$0.346~{\rm GeV}$}, 
\mbox{$\Delta_1=$} \mbox{$0.4~{\rm GeV}$}, and 
\mbox{$\mu_2=$} \mbox{$0.424~{\rm GeV}$},
\mbox{$\Delta_2=$} \mbox{$0.621~{\rm GeV}$}.
With this propagator, the lowest pseudo-threshold in the BSE is at
$2\mu_1 =$0.692~GeV.   The subsequent BSE solution is not to be 
taken as a physical represenation of the $\rho$ meson; its purpose
is to provide a test calculation where the propagator singularities
are within the domain of integation. 

\begin{figure}[t]
\includegraphics[width=3.0in,angle=-90]{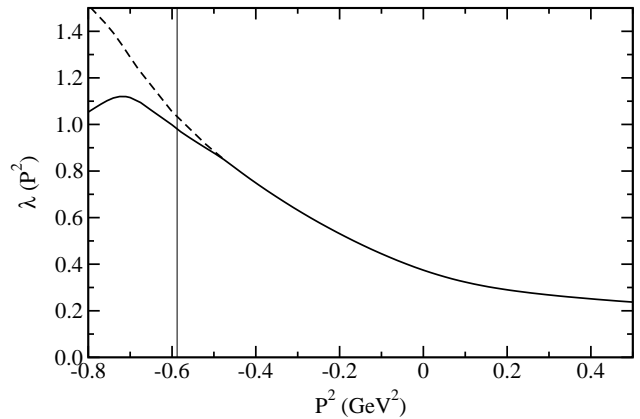}
\vspace*{-0.80in}
\caption{
The eigenvalue $\lambda(P^2)$ for the vector meson BSE.   
The mass-shell at \mbox{$\lambda(-m_{\rho}^2)=1$},  indicated by
the vertical line, is above the lowest pseudothreshold set by the
complex mass poles of the quark propagators.   Proper analytic continuation
of the integral produces the solid line; analytic continuation of
just the integrand produces the dashed line.  
\label{Fig:Lambda}  }
\end{figure}

The basis functions used to expand the $q^2$ dependence of the BS 
amplitudes are
\be
\phi_{s}(q^2) =\frac{N_s}{\Lambda} L_{s}(w) \; e^{-\frac{w^2}{2}}
\label{Def:phi}
\ee
where $w = q^2/\Lambda^2$, the typical hadron scale is 
$\Lambda =$ 0.800~GeV, $L_s(w)$ are Laguere polynomials, and $N_s$ is
fixed by the normalization condition given in \Eq{phi-orthog}.   
To obtain better than 1\% accuracy for the eigenvalue  $\lambda(P^2)$,
we found that 7 terms in this basis were sufficient.   The number of 
Chebyshev terms used for the angle basis was one.   The first
covariant from the set given in \Eq{vecBSAform} is known to be dominant
for the vector meson solution~\cite{Maris:1999nt} and this was 
the only one used here.  

The resulting BSE eigenvalue $\lambda(P^2)$ is given in \Fig{Fig:Lambda} as
the solid curve.
In the spacelike region ($P^2 > 0$) the eigenvalue is a monotonically,
decreasing function of $P^2$.  
When $P^2$ decreases below zero and becomes timelike, the eigenvalue
$\lambda(P^2)$ increases for a while then crosses unity, at 
$P^2=-M_{\rho}^2$, signifiying a bound state at $M_{\rho}= 0.770$~GeV.   
In \Fig{Fig:Lambda} this value of $P^2$ is shown as a thin, vertical
line.  On the timelike side of this line, the eigenvalue continues to
increase, then begins to decrease and ultimately goes to zero for 
($P^2 \ll 0$).   This is because the eigenvalue is a measure of the
magnitude of the BSE kernel and the propagators therein falloff as
the difference between the momentum arguments and the pole positions
becomes larger than any mass scale in the interaction.

The solid curve in \Fig{Fig:Lambda} represents the proper analytic 
continuation of the BSE from \mbox{$P^2>0$} to \mbox{$P^2<0$}.  The 
dashed line represents the result from a direct evaluation of the
BSE integral without regard to the possibility of singularities and
their movement as a function of $P^2$ in relation to the integration
domain.   That is, only a transcription from \mbox{$P^2>0$} to 
\mbox{$P^2<0$} has been made in the integrand of \Eq{Green2};
the Feynman technique for combining the denominators and shifting the
4-momentum variable to complete the square has not been made.  The
integration variables are $q^2$ and \mbox{$z_q=\hat{q} \cdot \hat{P}$}.  
It is clear that for the spacelike domain ($P^2 > 0$), and for the limited 
timelike domain for which no poles move into the domain of integration
(\mbox{$ -4\mu_1^2 < P^2 < 0$}, i.e., below the lowest pseudo-threshold), 
the two approaches give identical results.   The dashed line represents the 
approach to the BSE that has usually been implemented; it has  
been limited to states light enough to avoid the propagator singularities.

With the parameters given above, the quark poles are first 
encountered within the BSE integration when the analytic continuation reaches
\mbox{$P^2 = -4 \mu_1^2$}, that is,  at \mbox{$P^2 = -0.48~{\rm GeV}^2$}.  
Although the contibutions of the conjugate pair
of poles to the imaginary part of the BS eigenvalue $\lambda(P^2)$
cancel, their real contribution produces a discontinuity in the derivative 
$\lambda^\prime(P^2)$ at that point.  
Careful inspection of the solid curve in \Fig{Fig:Lambda} reveals that this
cusp occurs at the precise value where the full calculation and the
na\"{\i}ve calculation diverge.


\section{Discussion \label{Sec:Discuss}}

We have provided a method to obtain meson bound state solutions from the
Bethe-Salpeter equation in Euclidean metric when the dressed quark 
propagators have time-like complex conjugate mass poles within the 
integration domain.  This approximates features encountered in recent 
QCD modeling via the Dyson-Schwinger equations; the absence of real 
mass poles simulates quark confinement.   For this exploratory study
we represent the quark propagators as a sum of complex-conjugate mass 
poles, and project the BSE on to complete basis sets to represent 
the momentum and angle dependence.  We use Feynman integral techniques
to combine the propagator denominators and map the integration to a 
one-dimensional domain.  This allows a clear analysis of the analytic 
continuation in total meson momentum needed to reach the mass shell
and which causes the singularities to impinge upon the integration 
domain.   The BSE linear eigenvalue remains real; in other words the
eigenmass remains real.  The would-be decay width from one pole is exactly
cancelled by the effect of the partner pole; the meson is stable 
against decay
into a quark-antiquark pair.   This describes the confinement of quark and 
antiquark within the meson bound state, even though the meson mass is 
``above threshold''.

One of the limitations of the method presented here is that it relies
upon the projection of the BSE on to a complete basis set of functions
to represent the momentum dependence of the BS amplitudes.  In general the
behavior of the BS amplitudes is not known until after the BSE is 
solved.   If the chosen basis set is an inefficient representation,
one would expect this to result in a lack of convergence.   The 
(vector) meson mass that we have studied here is an integrated quantity
dominated by infrared physics;  it is not surprizing that convergence
is easily achieved with  the basis states of \Eq{Def:phi} which
fall off exponentially in the ultraviolet region.
However, we know from studies such as Ref.~\cite{Maris:1999nt} that the
leading ultraviolet power law behavior of the vector meson BS amplitudes 
is $1/q^2$.   Clearly the present basis would be inefficient for 
applications that depend strongly on the BS amplitude in this domain, such as 
the asymptotic behavior of the pion charge form factor~\cite{Maris:1998hc}. 
For such studies, appropriate basis sets would have to be utilized.

The principal reason that the BS amplitudes are expanded in terms of known 
basis functions is that the Feynman integral technique for handling
the two propagator denominators requires a complex variable shift for
the rest of the integrand.  In principle, the analytic continuation of 
the BS amplitude into the complex plane is determined by the dynamics 
of the BSE and can be determined only through the solution.  The 
implementation of the present method of solution requires a known 
analytic behavior for the BS amplitudes as expressed through the basis
functions.   One would expect an inadequacy of the basis in this respect to
show up as poor convergence.    

Future work using the  BSE solution method presented here will include
extension of the DSE modeling approach to SU(3) flavor meson states 
above 1 GeV and meson form factors for \mbox{$Q^2>3~{\rm GeV}^2$}.


\begin{acknowledgments}
This work is supported by the National Science Foundation under grant
Nos. PHY-0071361 and INT-0129236. 
\end{acknowledgments}
\bibliography{refsPM,refsPCT,refsCDR,refs,refsMAP}

\end{document}